# Piezostrain-induced in-plane anisotropy of $T_c$ for FeSe$_{0.5}$Te$_{0.5}$ thin films

Shu Mi[1,2,†], Ye Gao[1,2,†], Ze Jiang[3], Junwei Zhang[3], Wenhui Cao[4], Jinjin Li[4], Yutong Wang[1,2], Quan Liu[1,2], Chengchao Xu[5,6], Huaixin Yang[5,6], Jianqi Li[5,6], Guangming Zhang[1,2], Yayu Wang[1,2], Yonggang Zhao[1,2]*

[1]Department of Physics and State Key Laboratory of Low-Dimensional Quantum Physics, Tsinghua University, Beijing 100084, China

[2]Frontier Science Center for Quantum Information, Tsinghua University, Beijing 100084, China

[3]School of Materials and Energy, Electron Microscopy Centre of Lanzhou University and Key Laboratory of Magnetism and Magnetic Materials of the Ministry of Education, Lanzhou University, Lanzhou 730000, China

[4]National Institute of Metrology, Beijing 100029, China

[5]Beijing National Laboratory for Condensed Matter Physics and Institute of Physics, Chinese Academy of Sciences, Beijing 100190, China

[6]University of Chinese Academy of Sciences, Beijing 100049, China

[†]These authors contributed equally to this work.

*E-mail: ygzhao@tsinghua.edu.cn

**Abstract**

**Superconductivity is a macroscopic quantum phenomenon and it is commonly believed that the in-plane superconducting transition temperature ($T_c$) should be isotropic for superconductors with crystal structures of stacking layers (quasi two-dimensional). Here, we report the observation of piezostrain-induced in-plane anisotropy of $T_c$ in FeSe$_{0.5}$Te$_{0.5}$ (FST) thin films grown on ferroelectrics. The as-prepared FST shows tetragonal phase with equal in-plane lattice parameters and absence of in-plane anisotropy of $T_c$. Upon applying electric fields, piezostrain induces difference of the in-plane lattice parameters (*a-b*) and in-plane anisotropy of $T_c$. The in-plane anisotropy of $T_c$ correlates with *a-b* and becomes more remarkable for larger values of *a-b*. Some possible extrinsic effects were ruled out by experiments and analysis. It is suggested that electronic nematicity with one direction induced by piezostrain and inhomogeneous superconductivity in FST are involved in this unusual phenomenon. This work reveals the exotic behavior of FST and will stimulate exploration of similar behavior in other unconventional superconductors.**

Unlike the conventional superconductors, unconventional superconductors, such as the cuprate superconductors and iron-based superconductors, show exotic behaviors with nematicity, inhomogeneous superconductivity (inhomogeneous superfluid and gap), pair density waves, etc (*1-4*). These behaviors provide key ingredients for uncovering the mechanisms of unconventional superconductivity. Superconductivity is a macroscopic quantum phenomenon with dissipationless electrical current flow and perfect diamagnetism (Meissner effect) (*5*). The superconducting state can be expressed in terms of a macroscopic complex wave function with both amplitude and phase, i.e., all superconducting electrons (Cooper pairs) behaves like "one particle" with the same phase, involving spontaneous gauge symmetry breaking and Higgs phenomena (*6*), etc. In the general case, the superconducting transition temperature ($T_c$) should be isotropic for single-phase superconductors due to the macroscopic quantum nature. However, in some cases, anisotropy of $T_c$ were observed in superconductors with crystal structures of stacking layers (quasi two-dimensional) (*7*) or quasi one-dimensional chain (*8-10*), where coupling between layers or chains results in the $T_c$ of interlayer or interchain direction lower than that of the in-plane of layer or along the chain. It is believed that the in-plane $T_c$ should be isotropic for superconductors with crystal structures of stacking layers (quasi two-dimensional) and there has been no report on in-plane $T_c$ anisotropy so far. Iron-based superconductors $FeSe_{1-x}Te_x$ with crystal structures of stacking layers (quasi two-dimensional) have attracted enormous attention in the last decade for their simple lattice structure and abundant physical phenomena, such as a very high superconducting transition temperature ($T_c$) in the single-layer $FeSe/SrTiO_3$ (*11*) and topological superconductivity and Majorana bound states in $FeSe_{0.45}Te_{0.55}$ (*12,13*). Here we report the observation of piezostrain-induced in-plane $T_c$ anisotropy in $FeSe_{0.5}Te_{0.5}$ (FST) thin films grown on ferroelectric $Pb(Mg_{1/3}Nb_{2/3})_{0.7}Ti_{0.3}O_3$ (PMN-PT) single crystals. A possible mechanism is proposed to account for this unusual behavior by considering electronic nematicity with one direction induced by piezostrain and inhomogeneous superconductivity in FST.

FST films were fabricated on PMN-PT by a two-step growth method so as to optimize both crystallinity and superconducting property (*14*). The schematic configuration is displayed in Fig. 1A, with FST deposited at higher substrate temperature (HT-FST) and lower substrate temperature (LT-FST), respectively. The 25 nm insulating HT-FST film was deposited directly on (011)-oriented PMN-PT (fig. S1), which is beneficial to the sample crystallinity. Then 150 nm LT-FST film was deposited subsequently, which shows high quality with $T_c$ comparable to that of FST single crystal (*14*). Only FST (00*l*) and PMN-PT (0*ll*) peaks are present in the $\theta$-$2\theta$ scan (Fig. 1B), indicating the pure phase of the as-prepared FST with its *c*-axis perpendicular to the film surface. Due to the negligible HT-FST thickness compared to that of LT-FST, only peaks of LT-FST are found. High-angle annular dark-field scanning transmission electron microscopy (HAADF-STEM) was performed and demonstrates the good crystallinity of our sample (Fig.1C). The $\phi$ scan of the FST (101) peak and PMN-PT (101) peak show the in-plane four-fold symmetry of FST films and the in-plane two-fold symmetry of PMN-PT (Fig. 1D), consistent with the tetragonal phase of FST in spite of the orthogonal structure of the PMN-PT (011) crystal face. The tetragonal phase of the as-prepared FST is also demonstrated later by the measurements of the in-plane lattice parameters. The result of $\phi$ scan for the as-prepared FST indicates good in-plane alignment of FST. The in-plane orientation relationship between FST and PMN-PT (Fig. 1A) can be deduced from the $\phi$ scan as FST [100] || PMN-PT [100] and FST [010] || PMN-PT [01-1], suggesting the epitaxial orientation growth of FST films on PMN-PT

(*14, 15*).

The lattice distortion associated with the in-plane symmetry of FST film can be induced by electric fields, owing to the inverse piezoelectric effect of PMN-PT (011) (*16, 17*). Three typical strain states can be induced in PMN-PT by electric fields: as-prepared state (virgin state), -0 kV/cm state (with the out-of-plane electric field cycled between -4 kV/cm and +4 kV/cm for 5 times and then removed from -4 kV/cm) and +4 kV/cm state (under an electric field of +4 kV/cm). The schematic of polarization states and the resultant in-plane lattice evolution of PMN-PT for the three strain states is shown in Supplementary Information (fig. S2). Similar lattice evolution can be transformed into FST films due to the orientational epitaxial relationship between PMN-PT and FST. This has been verified by XRD peaks of FST (101) and FST (011) and the result is shown in Fig. 2A. The inset shows the in-plane lattice evolution of FST. After fitting the peak of FST (101), we can calculate the interplanar distance $d_{101}$ and thus the lattice constant $a$ from $1/d^2_{101} = (1/a^2) + (1/c^2)$, where $c$ can be derived by XRD patterns of $\theta$-$2\theta$ scan for FST (003) peak (Fig. 1B and fig. S3). The lattice constant $b$ can be obtained from FST (011) in the same way. For the as-prepared state, the peaks of FST (101) and FST (011) overlap, consistent with the tetragonal phase. For the -0 kV/cm state, the peaks of FST (101) and FST (011) separate with the former remaining unchanged and the latter shifting to the right, indicating an evident reduction of $b$ and no obvious change of $a$. The separation of the peaks reveals a transformation of FST from the tetragonal phase (as-prepared state) to an orthogonal phase (-0 kV/cm state). For the +4 kV/cm state, these two peaks get close to each other again, implying the decrease of *a* and the increase of *b*, making the orthogonality decline and approaching a tetragonal phase. The variations of *a* and *b* reveal an obvious in-plane anisotropy, especially for the -0 kV/cm state (Fig. 2B).

In order to investigate the effects of in-plane anisotropic strain on superconducting properties of FST, especially $T_c$ values related to the superconducting transitions along different directions, we designed and patterned FST with configuration along FST [100] and FST [010] at the same place (the inset of Fig. 3A). FST [100] and FST [010] are simplified as *a* [100] and *b* [010] hereafter. The temperature dependence of resistivity ($\rho$-$T$) was measured along *a* [100] and *b* [010] for the as-prepared, -0 and +4 kV/cm states, respectively (Fig. 3A). The $\rho$-$T$ behaviors along *a* [100] and *b* [010] overlap for the as-prepared state, then show a surprising separation for the -0 kV/cm state indicating in-plane anisotropy of $T_c$, and finally get close again for the +4 kV/cm state. The in-plane anisotropy of $T_c$ is quite surprising. To make sure the reliability and repeatability, we examined every step meticulously, employed different instruments and adopted different sequential orders for the measurements along *a* [100] or *b* [010]. The in-plane anisotropy of $T_c$ was repeated for different samples and the results of four more samples are presented as examples (fig. S4). Owing to the nonvolatile nature of the strain effect, the in-plane anisotropy of $T_c$ for the -0 kV/cm state can sustain (nonvolatile) as shown by the results of the same sample measured at different times (fig. S5).

The evolution of $T_c$ difference between *a* [100] and *b* [010] is consistent with that of the lattice constant difference between *a* and *b* (Fig. 2B), revealing the key role of the structural effects in the anisotropy of $T_c$. The values of $T_c^{onset}$ and $T_c^{zero}$ (defined in the caption) for different states are concluded in Fig. 3B. All $T_c$ values along both directions are elevated for the -0 kV/cm and +4 kV/cm states compared with those of the as-prepared state. The enhancement of $T_c$ can be attributed to the compression of the in-plane lattice, which has been reported for FST (*14, 18, 19*). The difference of $T_c^{onset}$ along *a* [100] and *b* [010] is not obvious except for the -0 kV/cm state, while a large anisotropy is observed in $T_c^{zero}$ especially for the -0 kV/cm state. Since the anisotropy of $T_c^{zero}$ is more remarkable, we will analyze superconducting properties by $T_c^{zero}$ hereafter.

In order to further analyze the relationship between lattice anisotropy and anisotropy of $T_c$, the lattice constants and $T_c$ under various electric fields between -0 and +4 kV/cm are measured (fig. S6). Variation of anisotropy of $T_c$ ($T_c^{zero}$ (*b*) – $T_c^{zero}$ (*a*)) with the in-plane lattice anisotropy *a-b* is shown in Fig. 3C. It can be seen that the anisotropy of $T_c$ correlates with lattice anisotropy *a-b*, and becomes more remarkable for larger *a-b*. The reversal of the ferroeletric domains near the coercive field $E_c$ is quite abrupt and unstable, resulting in the change of strains near $E_c$ unstable, so the data derived near $E_c$ (circled in Fig. 3C) have large measuring error.

It is well-known that superconductivity is a macroscopic quantum phenomenon, i.e., the sample becomes superconducting as a whole, so the existence of in-plane anisotropy of $T_c$ is quite surprising. To understand the mechanism behind this phenomenon, it's essential to rule out some artifacts, which may lead to the in-plane anisotropy of $T_c$.

First, microcracks may appear in PMN-PT after poled by electric fields and these microcracks may transfer to FST. If the orientations of cracks are regular along $a$ [100] and $b$ [010], it might lead to a $T_c$ anisotropy due to the different strains for microcracks along $a$ [100] and $b$ [010]. However, this probability has been excluded by the increasement of $T_c$ after poled (Fig. 3B) since cracks should lead to decrease of $T_c$ rather than increase. In addition, scanning electron microscopy (SEM) images for the -0 kV/cm state show no evident microcracks (fig. S7). Moreover, $T_c$ anisotropy has little correlation with the ferroelectric domains of PMN-PT (fig. S8), ruling out the contribution of the ferroelectric domains.

Another possibility is the regularly oriented grain boundaries along $a$ [100] and $b$ [010], leading to different superconducting transitions along these two directions. However, TEM results (fig. S9) demonstrate disordered grain boundaries. Moreover, as mentioned before, the $\phi$ scan of the FST (101) peak shows the in-plane four-fold symmetry (Fig. 1D) indicating good in-plane alignment of FST films with quasi-single crystal nature. This suggests that the grain boundaries are small-angle type, which has minor influence on electronic transport of the sample (*20*). It should be mentioned that the difference of $T_c^{onset}$ along $a$ [100] and $b$ [010] for the -0 kV/cm state is not consistent with the grain boundary scenario for which the $T_c^{onset}$ should be the same. Additionally, we carried out the same measurements on polycrystalline Nb films with irregular grain boundaries, and they exhibited the same $T_c$ along different directions after poled (fig. S10).

Therefore, the observed anisotropy of $T_c$ should be inherent in FST. Nematicity has been demonstrated in both cuprate superconductors (*1*) and iron-based superconductors (*2*) and it can break the in-plane four-fold symmetry leading to the anisotropy of transport along *a* [100] and *b* [010]. There are some reports on nematicity in FeSe$_{1-x}$Te$_x$ (*21-24*). In order to identify the nematicity in our samples, $\rho$-*T* behaviors in the whole temperature range are compared along the two perpendicular in-plane directions (*a* [100] and *b* [010]) for the as-prepared and -0 kV/cm state, respectively. The perfect superposition of $\rho$-*T* behaviors along *a* [100] and *b* [010] indicates a four-fold symmetry of resistivity in the as-prepared state (Fig. 4A). While for the -0 kV/cm state, $\rho$-*T* curves along *a* [100] and *b* [010] show evident differences especially at low temperatures (Fig. 4B) with resistivity along *b* [010] lower than that along *a* [100], especially below 50 K. The in-plane anisotropic resistivity has been observed in various kinds of superconductors, which is taken as a signature of the nematicity (*25-31*). In order to further explore the nature of nematicity, we calculated the elastoresistivity. The temperature dependence of elastoresistivity coefficient of B$_{1g}$ channel ($m_{11}$ - $m_{12}$) follows a Curie-Weiss dependence in a large temperature range (Fig. 4C), which has been taken as a direct evidence of the electronic nematicity in the literature (*26, 27, 29*), demonstrating electronic nematicity in our sample.

To understand the in-plane anisotropy of $T_c$, inhomogeneous superconductivity should also be considered in addition to the electronic nematicity. Strongly inhomogeneous superfluid (*3*) and spatial inhomogeneity of the superconducting gap (*4*) has been demonstrated in FeSe$_{1-x}$Te$_x$ in the nanoscale. These inhomogeneities are intrinsic and not caused by structural disorder resulting from the Se/Te alloying or local concentration of Se and Te atoms (*3, 4*). Moreover, crossover from the Bardeen-Cooper-Schrieffer (BCS) superconductivity to Bose-Einstein condensation (BEC) has been demonstrated in FeSe$_{1-x}$Te$_x$ (*32, 33*) and Josephson scanning tunnelling microscopy (STM) indicates that the Cooper pairs in FeTe$_{0.55}$Se$_{0.45}$ are very local: they are small in size and have little overlap compared with those in conventional superconductors (*3*). So nematicity, inhomogeneous superconductivity and BCS-BEC crossover with small size of Cooper pairs suggest dramatic difference of FST from the conventional superconductors (BCS superconductors). Considering the ultrashort superconducting coherence length (1.5 nm) of FeSe$_{0.5}$Te$_{0.5}$ (*34*), the inhomogeneous superconductivity results in coupling between nanoscale regions with high superfluid densities or large superconducting gaps (good superconducting regions) via nanoscale regions with low superfluid densities or small superconducting gaps (poor superconducting regions), and this coupling determines the $T_c^{zero}$. The in-plane anisotropic strain can lead to alignment of electronic nematicity and change of nematic fluctuation (*25, 35*), and it has been reported that the nematic fluctuation might have effects on superconductivity (*27, 28, 36-38*). Therefore, the in-plane anisotropic strain and electronic nematicity are expected to lead to different couplings between good superconducting regions for *a* [100] and *b* [010] of the -0 kV/cm state, resulting in the in-plane anisotropy of superconducting transition temperature. Based on this picture, it can be inferred that the resistant path along [110], which is the connection of *a* [100] and *b* [010] conducting paths, should be similar to that along *a* [100] because $T_c^{zero}$ is determined by the part with lower $T_c^{zero}$ in the path. To verify this, we measured *ρ-T* curves along [110] for the -0 kV/cm state. It is found that $T_c$ measured along [110] is same as that along *a* [100], but lower than that along *b* [010] (fig. S11).

The in-plane anisotropy of $T_c$ is quite unusual and more efforts are needed to under its mechanism. It is also interesting to explore similar behavior in other unconventional superconductors. In addition to the significance in fundamental issue, the in-plane anisotropy of $T_c$ may be also useful for applications. For example, the 109º ferroelastic domain switching in the (001)-oriented PMN-PT can rotate the in-plane uniaxial strain by 90º, and the nonvolatile in-plane uniaxial strain switches between two perpendicular directions by positive and negative electric-voltage pulses (*39*). So if the sample temperature is set around $T_c^{zero}$ of FST, one can switch the sample between the zero resistance state and nonzero resistance state by positive and negative electric-voltage pulses and the effect is nonvolatile.

**Materials and Methods**

Sample preparation

$FeSe_{0.5}Te_{0.5}$ stoichiometric target ($T_c$ =14 K) was prepared by solid state reaction method with powder materials of Fe (99.9%), Se (99.9%), and Te (99.999%). The well-mixed powders were cold pressed into discs, and then sealed in an evacuated ($10^{-2}$ Pa) quartz tube and heat treated at 300 °C for 10 h, then 600 °C for 40 h. The product was then reground and pressed into discs again, and heated at 650 °C for 40 h. $FeSe_{0.5}Te_{0.5}$ thin films were grown on the 0.5 mm-thick (011)-oriented PMN-PT substrates by Pulsed Laser Deposition (PLD) method. A KrF excimer (λ = 248 nm) at a repetition rate of 3 Hz was used, with a laser fluency 3 J/cm². 25 nm HT-FST was deposited at 450 ℃, and then 150 nm LT-FST was deposited at 300 ℃. The film thickness was measured by a Dektak 6M stylus profiler.

The instrument used for Nb film deposition was a DC magnetron sputtering system manufactured by Kurt J Lesker company, mainly consisting of a process chamber and a very small load-lock chamber. The cryo-pumped process chamber had a base pressure of about $1 \times 10^{-8}$ Torr and was equipped with several guns. The sputtering target was placed in the bottom of the chamber, while the substrate was placed in the middle of the top level. The substrate was cooled to about 17 ℃ by clamping to a water-cooled Cu substrate holder. The 3-inch Nb target inclined upward toward the center of the substrate. During sputtering, the substrate was kept at a uniform rotation for the purpose of obtaining a uniform film thickness. The target-to-substrate distance was set to 110 mm, and the sputtering target power was fixed at 500 W. The deposition rate was typically 0.75 nm/s at 5 mTorr, with an Ar flow of about 23 sccm.

Measurements

The crystal structure was characterized by a Rigaku SmartLab X-ray diffractometer with a Ge(220) × 2 incident beam monochromator, and the wavelength of Cu Kα1 is 1.5406 Å. The intersecting surface of FST was prepared by standard TEM sample preparation method for super-thin film via focused ion beam-scanning electron microscopy (FIB-SEM). The TEM samples were prepared using a cuprum membrane grid. The HAADF-STEM images were obtained using an FEI Titan Cs Probe transmission electron microscope, and the SAED patterns were obtained using an FEI

Titan Cryo Twin transmission electron microscope. Piezoresponse force microscopy (PFM) was carried out with the cantilever along the diagonal of the FE substrate. The $\pm 10$ V dc voltage biased by the tip on PMN-PT with a thickness of 0.5 mm can polarize the skin layer of the FE crystal because of the high electric field around the tip. SEM images were recorded with a Field Emission Scanning Electron Microscope (Zeiss Merlin). The transport properties were measured by four-probe configuration with a superconducting quantum interference device (MPMS 7 T, Quantum Design). The FST thin films were patterned with configuration along FST [100] and FST [010] at the same place with a size of 20 μm*100 μm (inset of Fig. 3A) through photo-lithography and ion-milling techniques.

**Acknowledgements**

We thank Jiun-Haw Chu, Shiliang Li and Wei Li for enlightening discussion on nematicity, as well as Weideng Sun and Wenhui Liang for helpful discussion on PMN-PT.

**Funding:** This work was supported by the Science Center of National Natural Science

Foundation of China (Grant No. 52388201), National Natural Science Foundation of China (Grant Nos. 52172270, 51831005, 11904195), the Basic and Applied Basic Research Major Programme of Guangdong Province, China (Grant No. 2021B0301030003), Jihua Laboratory (Project No. X210141TL210), National Key Research and Development Program of China under Grant 2018YFB2003401.

**Author contributions:** S.M. and Y.Z. planned the experiments. S.M. and Y.G. took part in FST sample preparation, property characterization and data analysis. Z.J. and J.Z. executed TEM experiments. W.C and Z.L grew Nb films and sample characterization. Y.W. and Q.L. performed the PFM measurements. C.X, H.Y. and J. L. prepared the FST target. The results were discussed with Y.W. and G.Z. and some key issues were clarified. The paper was written by S.M., Y.G. and Y.Z.. All authors reviewed and commented on the manuscript.

**Competing interests:** The authors declare no competing interests.

**Data availability:** The data that support the findings of this study are available from the corresponding author on reasonable request.

**Supplementary Material**

Supplementary Text

Figs. S1 to S11

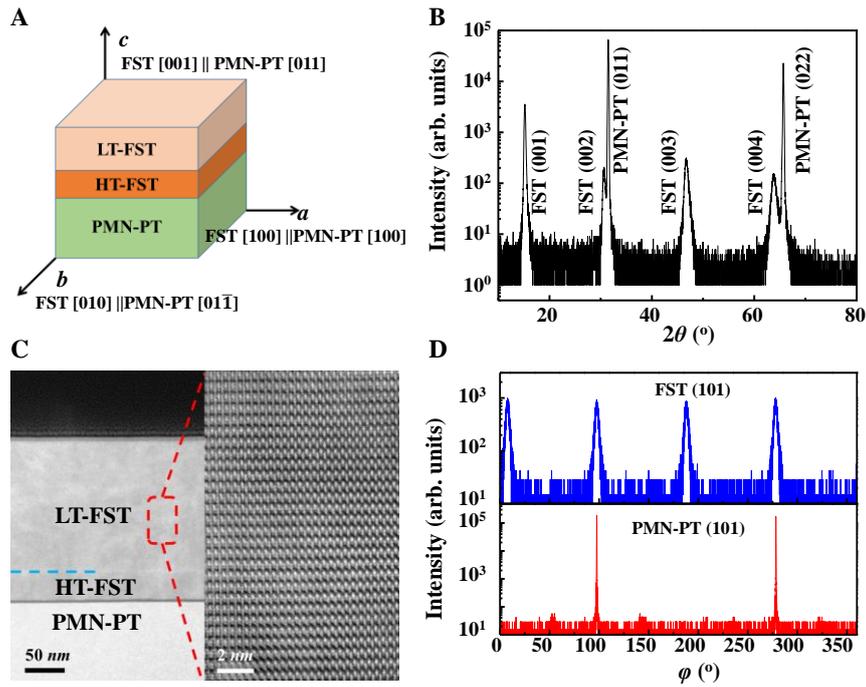

**Fig. 1. Structural properties. (A)** Schematic configuration of the sample, consisting of (011)-oriented PMN-PT substrate, 25 nm HT-FST and 150 nm LT-FST. **(B)** XRD patterns of $\theta$-$2\theta$ scan for our sample. HT-FST is too thin to be observed in XRD patterns. **(C)** Low magnification and high magnification cross-section views for the as-prepared films. LT-FST and HT-FST separated by a boundary shown by a blue dotted line. **(D)** $\phi$-scan results of FST (101) and PMN-PT (101).

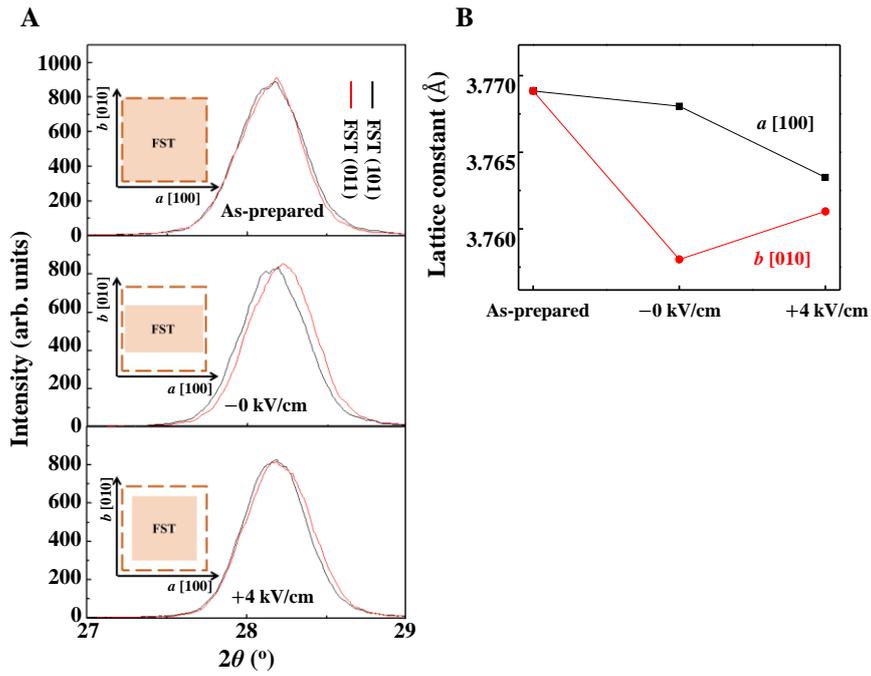

**Fig. 2. Structure evolution in different states.** (**A**) XRD patterns of θ-2θ scan for FST (101) (black lines) and FST (011) (red lines) for different states. The insets show the lattice evolution of FST, with the dashed orange squares representing the original sample geometry, and the solid orange squares demonstrating the sample deformation in different states. (**B**) Lattice constants *a* and *b* of FST calculated from peak positions for different states.

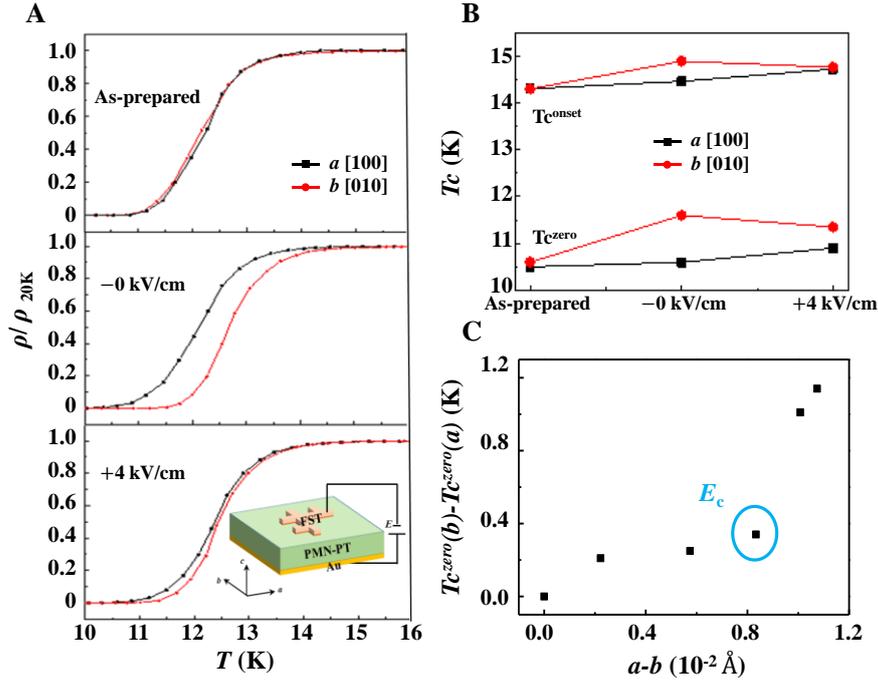

**Fig. 3. Effects of in-plane anisotropic strains on $T_c$.** **(A)** Temperature dependence of resistivity (normalized to that at 20 K) for *a* [100] (black lines) and *b* [010] (red lines) for different states. The inset is the schematic of the measurement configuration. Electric-field is applied from Au sputtered on the bottom to FST on the top. Hall bar is patterned along lattice *a* (FST [100]) and *b* (FST [010]) at the same place. **(B)** $T_c^{onset}$, and $T_c^{zero}$ for different states, where $T_c^{onset}$ is determined by the temperature at which $\rho$-$T$ curve deviates from the linear form in the normal state, and $T_c^{zero}$ is determined by the temperature at which resistivity reaches zero. **(C)** The relationship between $T_c$ anisotropy and lattice anisotropy.

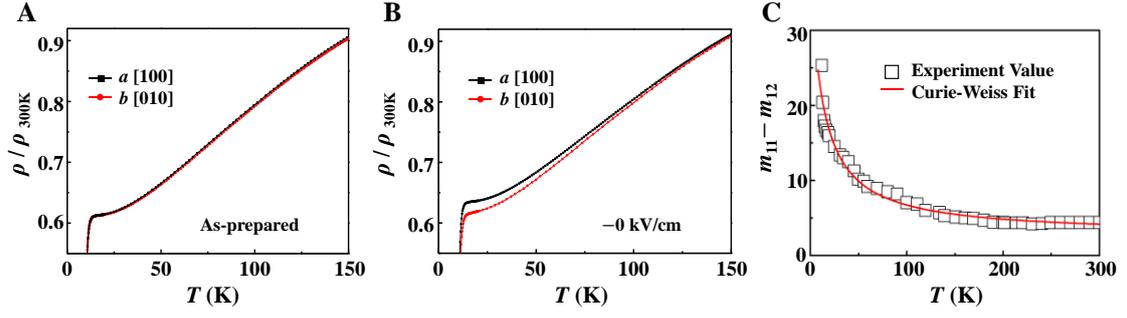

**Fig. 4. $\rho$-$T$ curves and nematicity. (A** and **B)** Comparisons of $\rho$-$T$ above $T_c$ measured along the two perpendicular in-plane directions for the as-prepared and -0 kV/cm states, respectively. Black and red lines represent $a$ [100] and $b$ [010]. **(C)** Temperature dependence of elastoresistivity coefficients $m_{11} - m_{12}$. Black squares denote experiment data, and red line represents Curie-Weiss fit above 14 K.